\documentclass[aps,pra,superscriptaddress,amsmath,amssymb,preprintnumbers,floatfix,showpacs,12pt]{revtex4}
\usepackage{amssymb}
\usepackage{epsfig}
\usepackage{subfigure}
\usepackage{graphicx}
\begin{document}
\title{ Variance squeezing and entanglement  of the $XX$ central spin model}

\author{Faisal A. A. El-Orany }
\email{el_orany@hotmail.com} \affiliation{ Department of
Mathematics and Computer Science, Faculty of Science, Suez Canal
University, Ismailia, Egypt; }

\author{M Sebawe Abdalla}

\affiliation{ Mathematics Department, College of Science, King
Saud University P.O. Box 2455, Riyadh 11451, Saudi Arabia }

\begin{abstract}

In this paper, we study the quantum properties for a system that
consists of a central atom interacting with surrounding spins
through the Heisenberg $XX$ couplings of equal strength. Employing
the Heisenberg equations of motion we manage to derive an exact
solution for the dynamical operators. We consider that the central
atom and its surroundings are initially prepared in the excited
state and in the coherent spin state, respectively. For this
system, we investigate the evolution of variance squeezing and
entanglement. The nonclassical effects have been remarked in the
behavior of all components of the system. The atomic variance can
exhibit revival–collapse phenomenon based on the value of the
detuning parameter.

\end{abstract}

\pacs{   03.65.Ge-42-50.Dv} \maketitle

\section{Introduction}
Quantum optics is based on the three fundamental processes,
namely, field-field interaction, atom-field interaction and
atom-atom interaction. The literature is quite rich by models
representing these processes. Recently, the interest in  the
atom-atom (, i.e., spin-spin) interactions is growing up as it is
a promising candidate  in implementing the quantum computer. In
this respect, we can find that  the demonstration of the  spin
dynamics in the semiconductor structures has been remarkably
increased in the last few years  in connection with the new
emerging areas of quantum computation and information
\cite{bayat}. There are various forms of the spin-spin systems
such as XX model \cite{lieb}, XY model \cite{jin}  and XXZ model
\cite{chiara}.

Spin models like Heisenberg spin chains or spin star systems,
describing for example a single electron spin in a semiconductor
quantum dot interacting with surrounding nuclear spins via
hyperfine coupling mechanisms, have been extensively studied
\cite{pratt}. These  models have been proved to be promising
candidates for the generation and the control of assigned quantum
correlations \cite{verstraete}. Additionally, the central spin
models can provide an appropriate description of quantum
information processes such as the quantum state transfer
\cite{[7]} and the quantum cloning \cite{[8]}.
 The Hamiltonian of the  $XX$ central spin model that is composed by a localized spin, hereafter
called central spin, coupled to $N$ spins with the same coupling
constant $\lambda$, takes the form
\cite{imamog,Privman,molmer,gaudin,molmeru}:

\begin{equation}\label{new1}
\hat{H}=\frac{\Omega}{2}\hat{S}_z+\frac{\mu}{2}\sum\limits_{j=1}^N\hat{\sigma}^j_z+\lambda
\sum\limits_{j=1}^N(\hat{S}_+\hat{\sigma}^j_-+\hat{S}_-\hat{\sigma}^j_+),
  \end{equation}
where $\hat{S}_i$ and $\hat{\sigma}^j_i (i=z,+,-)$  are the Pauli
operators referring to the central spin and the surrounding spins,
respectively; $\Omega$ and $\mu$  are the frequencies of the
central spin and the surrounding ones, where we assume that all
the surrounding spins have the same frequency. The architecture is
analogous to the star distribution networks used in
communications. Furthermore, this model is useful for sharing
entanglement between different spins \cite{molmeru}. In terms of
the collective operators, which describes the $N$ spins, the above
Hamiltonian can be expressed as
\cite{physica1,physica2,physica3,physica4,physr4}:
\begin{equation}
\frac{\hat{H}}{\hslash }=\frac{\Omega }{2}\hat{S}_{z}+\mu \hat{J}%
_{z}+\lambda \left(
\hat{J}_{+}\hat{S}_{-}+\hat{J}_{-}\hat{S}_{+}\right) , \label{1.7}
\end{equation}%
where $\hat{J}_{z}=\frac{1}{2}\sum\limits_{j=1}^N\hat{\sigma}^j_z$
and $\hat{J}_{\pm}=\sum\limits_{j=1}^N\hat{\sigma}^j_\pm$. These
operators satisfy the following commutation rules:
\begin{equation}
\left[ \hat{J}_{+},\hat{J}_{-}\right] =2\hat{J}_{z},\qquad \text{and}\qquad %
\left[ \hat{J}_{z},\hat{J}_{\pm }\right] =\pm \hat{J}_{\pm }.
\label{1.6}
\end{equation}
The Hamiltonian (\ref{new1}) has been frequently used in many
physical systems like quantum dots \cite{imamog}, two-dimensional
electron gases \cite{Privman} and optical lattices \cite{molmer}.
Additionally, it represents  a realization of the so-called Gaudin
model whose diagonalization has been derived via the Bethe ansatz
in \cite{gaudin}. This model has been frequently studied by
Messina and his coworkers
\cite{physica1,physica2,physica3,physica4}. The main objectives
  were to find the solution of the Schrodinger's equation or
to study the entanglement in the two-spin star system. In
\cite{physica1,physica2,physica3,physica4,physr4}  the authors
have considered the central star spin is the object and its
surroundings is the finite reservoir. Motivated by the importance
of this system in the literature we devote this paper for
demonstrating the quantum properties of the central spin and its
surroundings. Precisely, we investigate the variance squeezing and
the entanglement. Related to  the entanglement, we study the
entanglement between the central spin and the surroundings as well
as  among the spins of the surroundings. For the former we use the
linear entropy, while for the latter we adopt the concept of  spin
squeezing. From the analysis we show that the nonclassical effects
are noticeable in the behavior of all components of the system
for certain values of the interaction parameters. We proceed, for
reason will be clear in the following section we recast the
Hamiltonian (\ref{1.7}) in the following form:
\begin{equation}
\frac{\hat{H}}{\hslash }=\frac{\mu }{2}\hat{N}+\hat{C},
\label{2.3}
\end{equation}%
where
\begin{equation}
\hat{N}=\hat{J}_{z}+\frac{1}{2}\hat{S}_{z},\quad
\hat{C}=\frac{\Delta }{2}\hat{S}_{z}+\lambda \left( \hat{J}_{+}\hat{S}_{-}+%
\hat{J}_{-}\hat{S}_{+}\right) ,\qquad \Delta =\Omega -\frac{\mu
}{2}. \label{2.4}
\end{equation}
It is easy to prove that $\hat{N}$ and $\hat{C}$ are constants of
motion.

This paper is prepared  in the following order. In section  \textbf{%
II } we derive the basic relations of the system including the
solution of the Heisenberg equations of motion and the expectation
values of the dynamical operators.
 In section
\textbf{III}  we investigate the behavior of  the atomic
inversion. In section \textbf{IV} we discuss in details the
variance squeezing for the central spin  as well as for its
surroundings. In section \textbf{V} we study the entanglement. We
summarize our results  in section \textbf{VI}.

\section{Basic relations of the system}
The main goal of this section is to deduce the explicit form of
the dynamical operators of both the central spin $\hat{S}_i$ and
its surroundings $\hat{J}_i$, and also to evaluate the expectation
values for the different dynamical operators. Henceforth, we
denote the central spin and the surrounding spins by
$\widetilde{s}$ and $\widetilde{j}$, respectively.

As well known to study the temporal  behavior  of any quantum
system we have to find either the dynamical wave function or  the
dynamical operators.  For any dynamical operator $\hat{Q}$ the
Heisenberg equation of motion is given by
\begin{equation}
\frac{d\hat{Q}}{dt}=\frac{1}{i\hslash }\left[ \hat{Q},\hat{H}\right] +\frac{%
\partial \hat{Q}}{\partial t},  \label{2.1}
\end{equation}
where $\hat{H}$ is the Hamiltonian of the under consideration
system. By substituting  the Hamiltonian (\ref{1.7}) into the
equation (\ref{2.1}) for the different operators we arrive at:

\begin{eqnarray}
\frac{d\hat{J}_{z}}{dt} &=&i\lambda \left( \hat{J}_{-}\hat{S}_{+}-\hat{J}_{+}%
\hat{S}_{-}\right) ,\qquad \frac{d\hat{J}_{+}}{dt}=i\frac{\mu }{2}\hat{J}%
_{+}-2i\lambda \hat{J}_{z}\hat{S}_{+},  \nonumber \\
\frac{d\hat{J}_{-}}{dt} &=&-i\frac{\mu }{2}\hat{J}_{-}+2i\lambda \hat{J}_{z}%
\hat{S}_{-},\qquad \frac{d\hat{S}_{z}}{dt}=2i\lambda \left( \hat{J}_{+}\hat{S%
}_{-}-\hat{J}_{-}\hat{S}_{+}\right) ,  \nonumber \\
\frac{d\hat{S}_{+}}{dt} &=&i\Omega \hat{S}_{+}-i\lambda \hat{S}_{z}\hat{J}%
_{+},\qquad \frac{d\hat{S}_{-}}{dt}=-i\Omega \hat{S}_{-}+i\lambda \hat{S}_{z}%
\hat{J}_{-}.  \label{2.2}
\end{eqnarray}%
After performing some minor algebra on equations (\ref{2.2}) we
can obtain the following set of equations:
\begin{eqnarray}
\frac{d^{2}\hat{S}_{z}}{dt^{2}}+4\hat{C}^{2}\hat{S}_{z}=2\Delta \hat{C}%
,\qquad \frac{d^{2}\hat{J}_{z}}{dt^{2}}+4\hat{C}^{2}\hat{J}_{z}=4\hat{C}^{2}%
\hat{N}-\Delta \hat{C},\nonumber \\
\frac{d^{2}\hat{S}_{+}}{dt^{2}}-i\left( \mu +2\hat{C}\right) \frac{d\hat{S}%
_{+}}{dt}+\left[ 2\lambda ^{2}\left( \hat{N}-\frac{1}{2}\right) -\mu \hat{C}-%
\frac{\mu ^{2}}{4}\right] \hat{S}_{+} &=&0,  \nonumber \\
\frac{d^{2}\hat{S}_{-}}{dt^{2}}+i\left( \mu -2\hat{C}\right) \frac{d\hat{S}%
_{-}}{dt}-\left[ 2\lambda ^{2}\left( \hat{N}+\frac{1}{2}\right) -\mu \hat{C}+%
\frac{\mu ^{2}}{4}\right] \hat{S}_{-} &=&0,\nonumber \\
\frac{d^{2}\hat{J}_{+}}{dt^{2}}-i\left( \mu +2\hat{C}\right) \frac{d\hat{J}%
_{+}}{dt}+\left[ 2\lambda ^{2}\left( \hat{N}-\frac{1}{2}\right) -\mu \hat{C}-%
\frac{\mu ^{2}}{4}\right] \hat{J}_{+} &=&0,  \nonumber \\
\frac{d^{2}\hat{J}_{-}}{dt^{2}}+i\left( \mu -2\hat{C}\right) \frac{d\hat{J}%
_{-}}{dt}-\left[ 2\lambda ^{2}\left( \hat{N}+\frac{1}{2}\right) -\mu \hat{C}+%
\frac{\mu ^{2}}{4}\right] \hat{J}_{-} &=&0.  \label{2.5}
\end{eqnarray}%
Based on the fact  that $\hat{N}$ and $\hat{C}$ are constants of
motion, we can easily deduce the general solution of the equations
 (\ref{2.5}) as:
\begin{eqnarray}
\hat{S}_{z}(T) &=&\hat{S}_{z}\cos (2T\hat{D})+i\left( \hat{J}_{+}\hat{%
S}_{-}-\hat{J}_{-}\hat{S}_{+}\right) \frac{\sin (2T\hat{D})}{\hat{D}}%
+\bar{\Delta}\frac{\sin ^{2}(T\hat{D})}{\hat{D}^{2}}\hat{C},
\nonumber \\
\hat{S}_{+}(T) &=&\exp [i(\frac{\bar{\mu}}{2}+\frac{\hat{C}}{\lambda }%
)T]\left\{ \left[ \cos (T\hat{D}_{+})+i(\bar{\Delta}-\frac{\hat{C}}{%
\lambda })\frac{\sin (T\hat{D}_{+})}{\hat{D}_{+}}\right] \hat{S}%
_{+}\right.  \nonumber \\
&&-i\left. \frac{\sin (T\hat{D}_{+})}{\hat{D}_{+}}\hat{J}_{+}\hat{S}%
_{z}\right\} ,  \nonumber \\
\hat{S}_{-}(T) &=&\exp [-i(\frac{\bar{\mu}}{2}-\frac{\hat{C}}{\lambda }%
)T]\left\{ \left[ \cos (T\hat{D}_{-})-i(\bar{\Delta}+\frac{\hat{C}}{%
\lambda })\frac{\sin (T\hat{D}_{-})}{\hat{D}_{-}}\right] \hat{S}%
_{-}\right.   \nonumber \\
&&+i\left. \frac{\sin (T\hat{D}_{-})}{\hat{D}_{-}}\hat{J}_{-}\hat{S}%
_{z}\right\},\nonumber\\
\hat{J}_{z}(T) &=&\hat{J}_{z}\cos (2T\hat{D})-i\left( \hat{J}_{+}\hat{%
S}_{-}-\hat{J}_{-}\hat{S}_{+}\right) \frac{\sin (2T\hat{D})}{2\hat{D}%
}+\left( 2\hat{C}\hat{N}-\frac{\bar{\Delta}}{2}\right) \frac{\sin ^{2}(T\hat{%
D})}{\hat{D}^{2}}\hat{C},  \nonumber \\
\hat{J}_{+}(T) &=&\exp [i(\frac{\bar{\mu}}{2}+\frac{\hat{C}}{\lambda }%
)T]\left\{ \left[ \cos (T\hat{D}_{+})-i\frac{\hat{C}}{\lambda
}\frac{\sin
(T\hat{D}_{+})}{\hat{D}_{+}}\right] \hat{J}_{+}\right.  \nonumber \\
&&-2i\left. \frac{\sin (T\hat{D}_{+})}{\hat{D}_{+}}\hat{J}_{z}\hat{S}%
_{+}\right\} ,  \nonumber \\
\hat{J}_{-}(T) &=&\exp [-i(\frac{\bar{\mu}}{2}-\frac{\hat{C}}{\lambda }%
)T]\left\{ \left[ \cos (T\hat{D}_{-})-i\frac{\hat{C}}{\lambda
}\frac{\sin
(T\hat{D}_{-})}{\hat{D}_{-}}\right] \hat{J}_{-}\right.   \nonumber \\
&&+2i\left. \frac{\sin (T\hat{D}_{-})}{\hat{D}_{-}}\hat{J}_{z}\hat{S}%
_{-}\right\},
 \label{2.6}
\end{eqnarray}
where we have defined%
\begin{equation}
\hat{D}=\sqrt{\frac{\bar{\Delta}^{2}}{4}+\hat{J}^{2}-\hat{N}^{2}+\frac{1%
}{4}},\qquad \hat{D}_{\pm }=\sqrt{\frac{\bar{\Delta}^{2}}{4}+\hat{J}^{2}+%
\frac{1}{4}-(\hat{N}\pm 1)^{2}}  \label{2.11}
\end{equation}%
and the normalized quantities $\bar{\Delta}=\Delta /\lambda
,\bar{\mu}=\mu /\lambda $ and $T=\lambda t$. Throughout the
solution,  for any operator, say, $\hat{R}(T)$ at the initial time
we have used the short-hand notation $\hat{R}(0)=\hat{R}$. As far
as we know this the first time the solution of  the Heisenberg
equations of the system (\ref{2.3}) to be presented.

Having obtained  the dynamical operators we are therefore in a
position to discuss some statistical properties of the system. To
this end, we  evaluate the different expectation values of the
above dynamical operators. For doing so, we assume that the
$\widetilde{s}$ system is initially in its excited state
$|+\rangle $, while the $\widetilde{j}$ system is initially in the
atomic coherent state or a coherent spin state (CSS) $|\theta
,\phi \rangle $ \cite{spin1}. The CSS is a minimum uncertainty
state that describes a system with a well-defined relative phase
between the species \cite{choi}. The CSS may be written in the
form:
\begin{equation}
|\theta ,\phi \rangle =\sum\limits_{m=-j}^{j}R_{m}^{j}(\theta
,\phi )|j,m\rangle ,  \label{3.1}
\end{equation}%
where
\begin{equation}
R_{m}^{j}(\theta ,\phi )=\sqrt{ C_{j+m}^{2j} }\exp [i(j-m)\phi
]\cos ^{j+m}(\frac{\theta }{2})\sin ^{j-m}(\frac{\theta }{2})
\label{3.2}
\end{equation}%
and $C_{m}^{n}$ is the binomial coefficient. Thus, the initial
factorized state of the  system is $|\psi \rangle =|\theta ,\phi
\rangle \otimes |+\rangle $. In the following calculations  we use
the slowly
varying operators, i.e. $\hat{S}_\pm(T)\equiv \hat{S}_\pm(T) \exp (\mp i\frac{\bar{\mu}}{2}%
T),\quad \hat{J}_\pm(T)\equiv \hat{J}_\pm(T) \exp (\mp i\frac{\bar{\mu}}{2}%
T)$. The choice of the slowly varying operators adopted to
simplify the treatment of the system.  With this in mind and using
the solution (\ref{2.6}) we can easily calculate the different
moments as:
\begin{widetext}
\begin{eqnarray}\label{insert2}
\begin{array}{lr}
\langle \hat{S}_{z}(T)\rangle
=\sum\limits_{m=-j}^{j}|R_{m}^{j}(\theta ,\phi
)|^{2}\left[ \cos (2Tg(m))+\frac{\bar{\Delta}^{2}}{2g^{2}(m)}\sin ^{2}(Tg(m))%
\right],
\\
\\
\langle \hat{S}_{+}(T)\rangle =-i
\sum\limits_{m=-j}^{j-1}R_{m}^{j}(\theta ,\phi )R_{m+1}^{\ast
j}(\theta ,\phi )    \frac{\Lambda (m)}{2g(m)}\left[ \sin
(2Tg(m))+i\frac{\bar{\Delta}}{g(m)}\sin ^{2}(Tg(m))\right],
\\
\\
\langle \hat{S}_{-}(T)\hat{S}_{+}(T)\rangle
=\sum\limits_{m=-j}^{j}|R_{m}^{j}(\theta ,\phi )|^{2}\Lambda ^{2}(m)\frac{%
\sin ^{2}(Tg(m))}{g^{2}(m)},\\
\\
\langle \hat{J}_{z}(T)\rangle
=\sum\limits_{m=-j}^{j}|R_{m}^{j}(\theta
,\phi )|^{2}\left\{ m\cos (2Tg(m))+\left[ 2m\left( \frac{\bar{\Delta}^{2}}{4}%
+\Lambda ^{2}(m)\right) +\Lambda ^{2}(m)\right] \frac{\sin ^{2}(Tg(m))}{%
g^{2}(m)}\right\},\\
\\
\langle \hat{J}_{+}(T)\rangle =\sum%
\limits_{m=-j}^{j-1}R_{m}^{j}(\theta ,\phi )R_{m+1}^{\ast j}(\theta ,\phi )%
\frac{K(m)}{\Lambda (m)},\\
\\
\langle \hat{J}_{-}(T)\hat{J}_{+}(T)\rangle
=\sum\limits_{m=-j}^{j}|R_{m}^{j}(\theta ,\phi )|^{2}\Lambda
^{2}(m)\left[ 1-2\left( m+1\right) \frac{\sin
^{2}(Tg(m))}{g^{2}(m)}\right],\\
\\
\langle \hat{J}_{+}^{2}(T)\rangle =\sum%
\limits_{m=-j}^{j-2}R_{m}^{j}(\theta ,\phi )R_{m+2}^{\ast
j}(\theta ,\phi
)\left\{ K(m)K(m+1)\right.
\\
 \\
+G(m)\left[ h_{1}(T,m+1)\Lambda ^{2}(m+2)h_{2}(T,m+2) - \left.
\Lambda ^{2}(m+1)h_{1}(T,m+2)\right] \right\},
\end{array}
\end{eqnarray}
\end{widetext}
where
\begin{widetext}
\begin{eqnarray}\label{insert22}
\begin{array}{lr}
g(m)=\sqrt{\frac{\bar{\Delta}^{2}}{4}+\Lambda ^{2}(m)},\qquad \Lambda (m)=%
\sqrt{\left( j-m\right) \left( j+m+1\right) } ,
\\
\\
h_{1}(T,m)= \cos (Tg(m))+i\frac{\bar{\Delta}}{2}\frac{\sin (Tg(m))}{%
g(m)} ,\qquad h_{2}(T,m)=\frac{\sin (Tg(m))}{g(m)},  \\
\\
G(m) =i\Lambda (m)\Lambda (m+1)[h_{1}^{\ast
}(T,m)h_{2}(T,m+1)-h_{2}(T,m)h_{1}^{\ast }(T,m+1)],\\
\\
\frac{K(m)}{\Lambda (m)} =\left[ h_{1}^{\ast
}(T,m)h_{1}(T,m+1)+\Lambda ^{2}(m+1)h_{2}(T,m)h_{2}(T,m+1)\right]
\end{array}
\end{eqnarray}
\end{widetext}
and $\langle \hat{S}_{-}(T)\rangle =\langle \hat{S}_{+}(T)\rangle
^{\ast }, \quad \langle \hat{J}_{-}(T)\rangle =\langle \hat{%
J}_{+}(T)\rangle ^{\ast } .$

In the following sections we investigate the quantum properties of
both the $\widetilde{s}$ and $\widetilde{j}$ systems using the
results of the current section. The investigation will be confined
to the atomic inversion, variance squeezing, linear entropy and
spin squeezing. Throughout the investigation we always consider
$\phi=0$ for the sake of simplicity.

%%%%%%%%%%%%%%%%%%%%%%%%%%%%%%%%%%%%%%%%%%%%%%%%%%%%%%%%%%
\section{ Atomic inversion}
%%%%%%%%%%%%%%%%%%%%%%%%%%%%%%%%%%%%%%%%%%%%%%%%%%%%%%%%%%
Atomic inversion represents an important  physical quantity, which
is defined as the difference between the probabilities of finding
the atom in the exited state and in the ground state.
Investigating  the behavior  of the atomic inversion can provide a
lot of information about the system. For instance, the atomic
inversion of the standard Jaynes-Cummings model (JCM) \cite{1} is
well known in the quantum optics by exhibiting  the
revival-collapse phenomenon (RCP), which  reflects the nature of
the statistics of the radiation field. The features of the RCP of
the JCM have been analyzed in details in \cite{eber}. Moreover, it
has been shown that the envelope of each revival is a readout of
the photon distribution, in particular, for the states whose
photon-number distributions are slowly varying \cite{fle}. For the
system under consideration, i.e. the $XX$ central spin model,
we'll show that this fact is not always applicable.
 Therefore, we devote this section to
discuss the atomic inversion of the central atom, i.e. $\langle
\hat{S}_{z}(T)\rangle$, which has been  calculated  in the
previous section. It is worth reminding that $\hat{N}$ is a
constant of motion and hence it is not necessary   to present the
quantity $\langle \hat{J}_{z}(T)\rangle$. Throughout the numerical
investigation in this paper  we confine ourselves to $j=150$ and
$\theta=\pi/2,\quad \pi/3$.  For these two cases of $\theta$ the
photon-number distribution of the CSS has smooth envelope, which
is symmetric around the origin only for  $\theta=\pi/2$.
 We proceed, for the atomic inversion we have plotted
figures (1) and (2)  for different values of $j,\bar{\Delta}$ and
$\theta $. %%%%%%%%%%%%%%%%%%%%%%%%%%%%%%%%%%%%%%%%%%%%%%%%%%%%%%%%%%
\begin{figure}[tbh]
\centering
\includegraphics[width=35.pc,height=15pc]{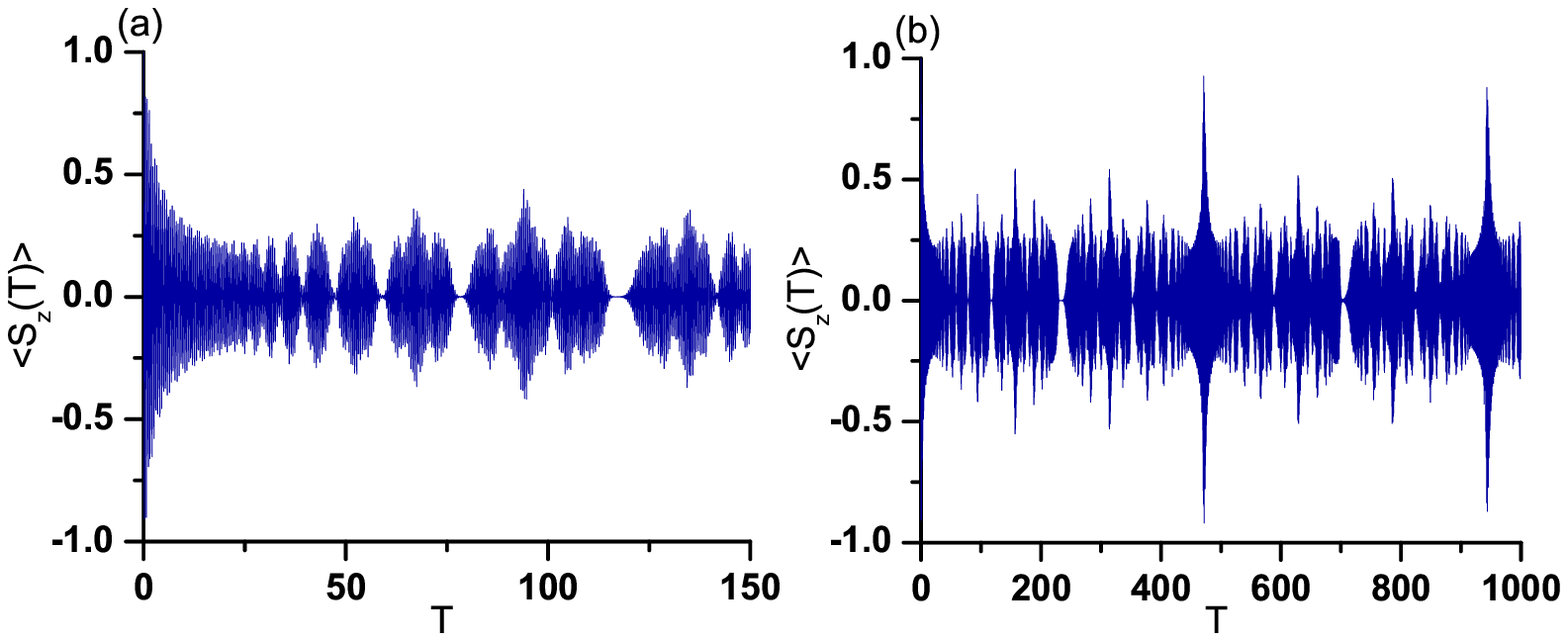}
\caption{The atomic inversion $\langle S_{z}\left( T\right)
\rangle $
against the scaled time $T$ for
$(j,\protect\theta,\bar{\Delta})=(150,\protect\pi %
/2,0 )$.}\end{figure}

\begin{figure}[tbh]
\centering
\includegraphics[width=35.pc,height=15pc]{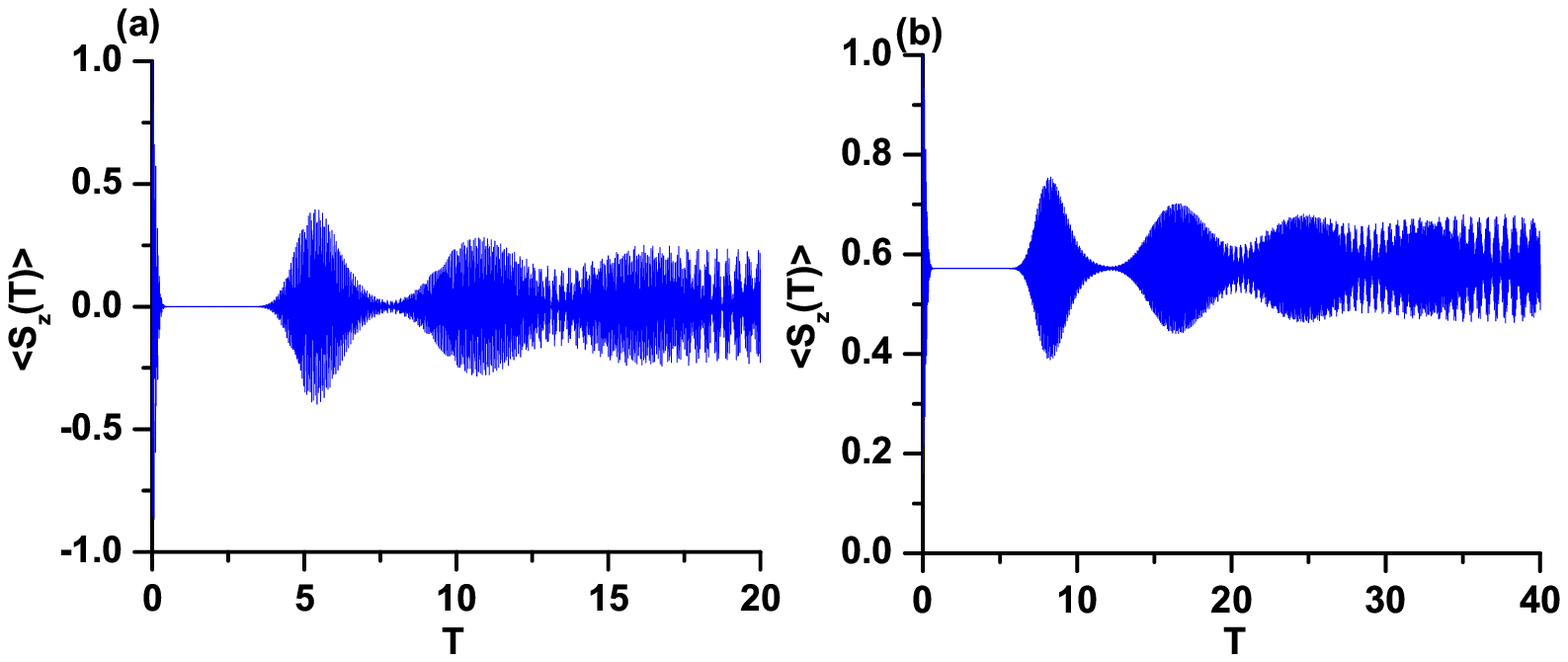}
\caption{The atomic inversion $\langle S_{z}\left( T\right)
\rangle $
against the scaled time $T$  for $(j,\protect\theta)=(150,\protect\pi %
/3 )$ with $\bar{\Delta}=0,$ (a) and $300$ (b). }
\end{figure}
%%%%%%%%%%%%%%%%%%%%%%%%%%%%%%%%%%%%%%%%%%%%%%%%%%%%%%%%%%%%%%%%%%%%%%%%%%%%%%%%%
We have considered in Figs. 1 the case in which $j=150,$ $%
\bar{\Delta}=0$ and $\theta =\pi /2$.  Fig. 1(a) displays the
behavior of the function against the time within the interval
$T\in [0,150]$. From this figure   we can observe long series of
the particular type of the revival patterns with rapid
fluctuations around the origin. Furthermore,  the collapse periods
increase as the interaction time increases, too. For a large
 range of the interaction time $T$, $\langle S_{z}\left( T\right)
\rangle $ exhibits more fluctuations and interference between the
patterns, which causes   the superstructure phenomenon (see Fig.
1(b)). From this figure it is obvious that $\langle
\hat{S}_{z}(T)\rangle$ is a periodic function with period $\tau
\simeq 480$. This behavior is completely different from that of
the JCM \cite{eber}. In the language of entanglement, the systems
$\widetilde{s}$ and $\widetilde{j}$ are periodically disentangled.
We'll get back to this point when we discuss the linear entropy.
Now we draw the attention to Figs. 2, where $\theta =\pi /3,
j=150$ and $\bar{\Delta}=0, 300$. From these figures we can
observe the standard RCP, which is quite similar to that of the
JCM. The origin in this is the shape of the photon-number
distribution of the CSS, which, in this case, is quite similar to
that of the coherent state. Moreover, the comparison between Fig.
1(b) and Fig. 2(a) shows that the smoothness  of  the photon
distribution does not always cause  the standard RCP. Precisely,
this phenomenon, for the system under consideration, depends on
the location of the center of symmetrization of the photon-number
distribution in the photon-number domain. As a result of this the
behavior of the system for $\theta=\pi/2$ is quite different from
that of the JCM \cite{fle}. We proceed, from Fig. 2(a) we can
observe long period of collapse after the initial revival pattern.
This means that the correlation between the bipartite (, i.e. the
$\widetilde{s}$ and $\widetilde{j}$ systems) is weak  during this
period of the interaction. It is worth reminding that the JCM  can
generate Schr\"{o}dinger-cat states in the course of this collapse
period \cite{cat}. Nevertheless, it is difficult to justify this
fact for our system.  Afterward, the $\langle S_{z}\left( T\right)
\rangle $ exhibits a series of the revival patterns, which
interfere with each others showing a chaotic behavior.
 When the effect  of the detuning parameter is considered in the interaction,
 $\bar{\Delta}=300$,  the atomic inversion is shifted up  and
 shown
 fluctuations   around $0.6.$ Also we can observe an increase
in the revival-pattern series compared to that of the exact
resonance case. The shift occurred  in the evolution  of the
$\langle \hat{S}_{z}(T)\rangle$ for the off-resonance case
indicates weak entanglement in the bipartite. This point will be
elaborated when we discuss the behavior of the linear entropy.

\section{Variance squeezing}

One of the most important  phenomena in the  quantum optics is the
squeezing, which has been attracted much attention in the last
three decades. Thereby, it is reasonable to discuss this
phenomenon for the  $XX$ central spin model. Precisely, we
investigate squeezing for both of the $\widetilde{s}$ and
$\widetilde{j}$ systems. In doing so, we assume that $\hat{A}$ and
$\hat{B}$ are any two physical observable, which satisfy the
commutation relation $\left[ \hat{A},\hat{B}\right] =i\hat{C}$.
Thus, we have the following uncertainty  relation:
\begin{equation}
\langle \left( \Delta \hat{A}\right) ^{2}\rangle \langle \left( \Delta \hat{B%
}\right) ^{2}\rangle \geq \frac{1}{4}|\langle \hat{C}\rangle|^{2},
\label{4.1}
\end{equation}%
where $\langle \left( \Delta \hat{A}\right) ^{2}\rangle =\langle \hat{A}%
^{2}\rangle -\langle \hat{A}\rangle ^{2}$.
 Therefore, squeezing  can be generated   in the component
$\hat{A}$, say,  if  the following inequality is  satisfied:
\begin{equation}
A = \langle \left( \Delta \hat{A}\right)
^{2}\rangle-\frac{|\langle \hat{C}\rangle| }{2} <0. \label{4.2}
\end{equation}%
Similar inequality can be quoted to the operator $\hat{B}$. As
well-known, the atomic squeezing is  state dependent. If we use
different form of the commutation rules, we will get different
behavior in the evolution of the squeezing factors. Based on this
fact it would be more convenient to investigate  these various
cases. Precisely, we study squeezing for the cases $(x,y), (x,z)$
and $(y,z)$. For instance, for the $(x,y)$ case of the
$\widetilde{s}$ system we mean
 $\hat{A}=\hat{S}%
_{x},\hat{B}=\hat{S}_{y}$ and $\hat{C}=\hat{S}_{z}$. The other
cases can be similarly expressed.  Information about these cases
have been shown in Figs. 3--5. We start the discussion with Figs.
3, which have been given to the case $(x,y)$. Generally, we have
noted that for the resonance case squeezing can occur only in the
$y$-component. This is remarkable from the expressions
(\ref{insert2}), where  $\langle\hat{S}%
_{x}(T)\rangle=0$ and hence $S%
_{x}>0$. This indicates  that the value of the detuning parameter
$\bar{\Delta}$ plays a significant role in the behavior of the
atomic squeezing. From Fig. 3(a), i.e.  resonance case, one can
observe that squeezing occurs only in $S_{y}$ and shows maximum
after onset the interaction. As the interaction time increases the
amount of squeezing smoothly decreases. Also, the function shows
irregular fluctuations. Quite different behavior is observed for
the off-resonance case as shown in Fig. 3(b).  In this case, the
quantity $S_{y}$ exhibits amount of squeezing less than that of
the resonance case. Surprisingly, it also exhibits RCP. This is
related to the fact that the Rabi oscillation is quite sensitive
to the value of $\bar{\Delta}$ (cf. (\ref{insert22})). This
phenomenon has been reported earlier in the evolution of the
squeezing factors of the Kerr coupler \cite{fai}. %%%%%%%%%%%%%%%%%%%%%%%%%%%%%%%%%%%%%%%
%%%%%%%%%%%%%%%%%%%%%%%%%%%%%%%%%%%%%%%%%%%%%%%%%%%%%%%%%%%%%%%%%%%%%%%%%%%%%
\begin{figure}[tbh]
\centering
\includegraphics[width=35.pc,height=30pc]{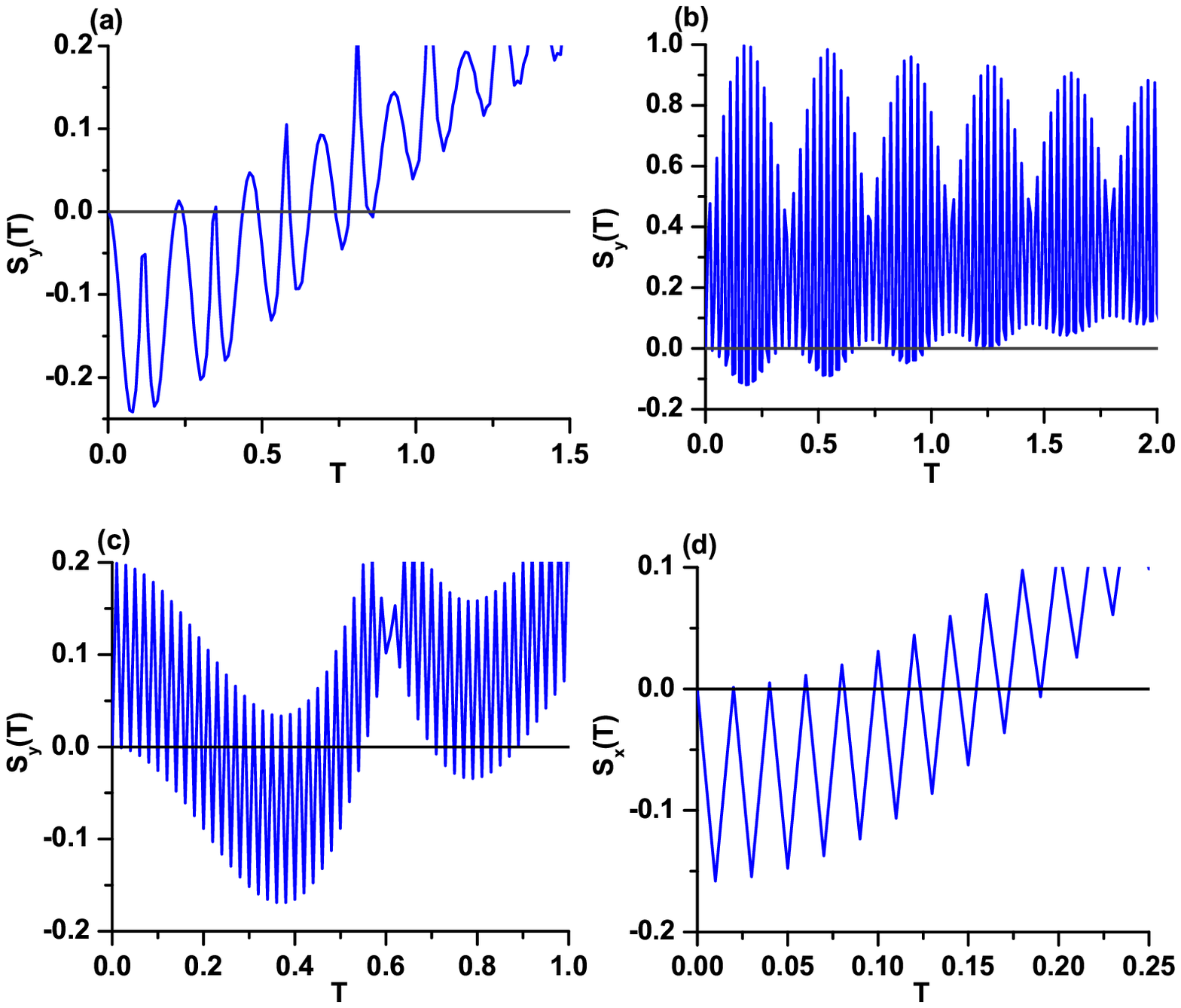}
\caption{The variances squeezing as indicated for the case $(x,y)$
against the scaled time $T$  when $(j,\protect\theta)=(150,\protect\pi %
/2)$ with  $\bar{\Delta}=0$ (a),  $300$ (b), and $100$  (c)-(d).
The straight line represents the squeezing bound. }
\end{figure}
%%%%%%%%%%%%%%%%%%%%%%%%%%%%%%%%%%%%%%%%%%%%%%%%%%%%%%%%%%%%%%%%%%%%%%%%
%%%%%%%%%%%%%%%%%%%%%%%%%%%%%%%%%%%%%%%%%%%%%%%%%%%%%%%%%%%%%%%
 Figs. 3(c) and (d) shed some light on the important role of the $\bar{\Delta}$.
From these figures we can observe the occurrence of squeezing in
the two components $x$ and $y$.  In the $y$-component,  squeezing
occurs over a range of interaction time broader than that in the
$x$-component. Nonetheless, the variance squeezing of the
$x$-component  displays the maximal squeezing after onset  the
interaction. The comparison among  Figs. 3(a)-(d) confirms the
potential for controlling the quantum  properties of the system
via externally adjustable parameters such as duration of
interaction time, detuning, etc. Finally, for $\theta=\pi/3$ we
have found that  squeezing in the case $(x,y)$ is almost
negligible.

%%%%%%%%%%%%%%%%%%%%%%%%%%%%%%%%%%%%%%%%%%%%%%%%%%%%%%%%%%%%%%%%%%
Now we draw the attention to Figs. 4 (with $\theta=\pi/2$) and 5
(with $\theta=\pi/3$), which have been given to the cases $(x,z)$
and $(y,z)$. Generally, we have found that the amount of squeezing
in the $z$-component is almost negligible. Moreover, for these
cases the system can generate squeezing for  interaction time
larger than that of  the $(x,y)$  case.  In Figs. 4(a) and 4(b) we
present the case $(x,z)$. From these figures,  we can observe that
squeezing occurs in the $x$-component only after onset  the
interaction.  As the interaction time increases the amount of the
squeezing gradually decreases. For the off-resonance case, we can
observe the occurrence of the RCP in the evolution of the
squeezing function (see Fig. 4(b)).  Fig. 4(c) is given for the
case $(y,z)$ with strong detuning.  From this figure we can
observe the occurrence of the long-lived squeezing. Moreover, more
fluctuations can be seen in addition to interference between the
patterns showing  the superstructure phenomenon. Now, we draw the
attention to the case  $\theta =\pi /3$, which is shown in Figs.
5. For the case $(x,z)$ squeezing occurs only in the resonance
case, while for  the case $(y,z)$ it occurs in the off-resonance
case, i.e. $\bar{\Delta}=300.$ In Fig. 5(a), squeezing is
immediately generated
 after switching on the interaction showing its maximal
value. As the interaction proceeds squeezing appears in an
oscillatory form, which decreases with increasing the interaction
time. This behavior is quite similar to that of the single-atom
entropy squeezing of the two two-level atoms interacting with a
single-mode quantized electromagnetic field in a lossless resonant
cavity \cite{fais}.
  The comparison
between Fig. 5(a) and Fig. 5(b) shows that  squeezing observed for
the case $(x,z)$ is stronger than that for  the case $(y,z)$. From
Figs. (3)-(5) one can conclude that squeezing can be engineered (,
i.e. it can be obtained from a specific component) via controlling
the interaction parameters.

%%%%%%%%%%%%%%%%%%%%%%%%%%%%%%%%%%%%%%%%%%%%%%%%
\begin{figure}[tbh]
\centering
\includegraphics[width=35.pc,height=15pc]{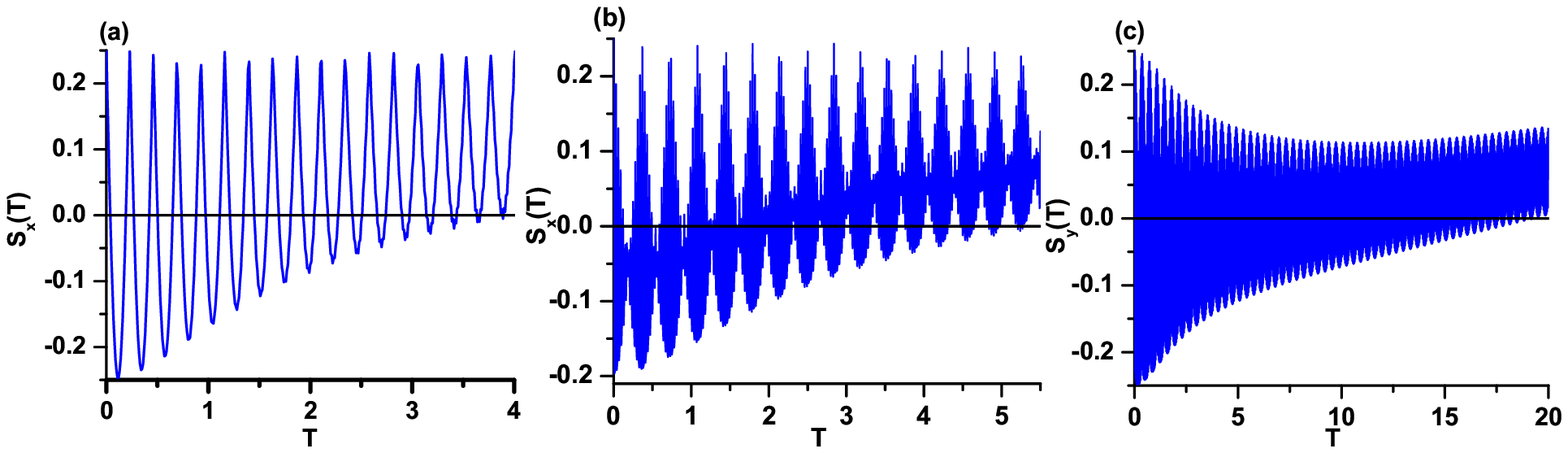}
\caption{ The variances squeezing as indicated
against the scaled time $T$  for $(j,\protect\theta)=(150,\protect\pi %
/2)$ with  $\bar{\Delta}=0$ (a),  $300$ (b)--(c). Figures (a) and
(b) are given to the case $(x, z)$, while (c) for the case $(y,
z)$. The straight line represents the squeezing bound. }
\end{figure}
%%%%%%%%%%%%%%%%%%%%%%%%%%%%%%%%%%%%%%%%%%%%%%%%%%%%%%%%%%

%%%%%%%%%%%%%%%%%%%%%%%%%%%%%%%%%%%%%%%%%%%%%%%%%%%%%%%%
%%%%%%%%%%%%%%%%%%%%%%%%%%%%%%%%%%%%%%%%%%%%%%%%%%%%%%%%%%%%%
\begin{figure}[tbh]
\centering
\includegraphics[width=30.pc,height=15pc]{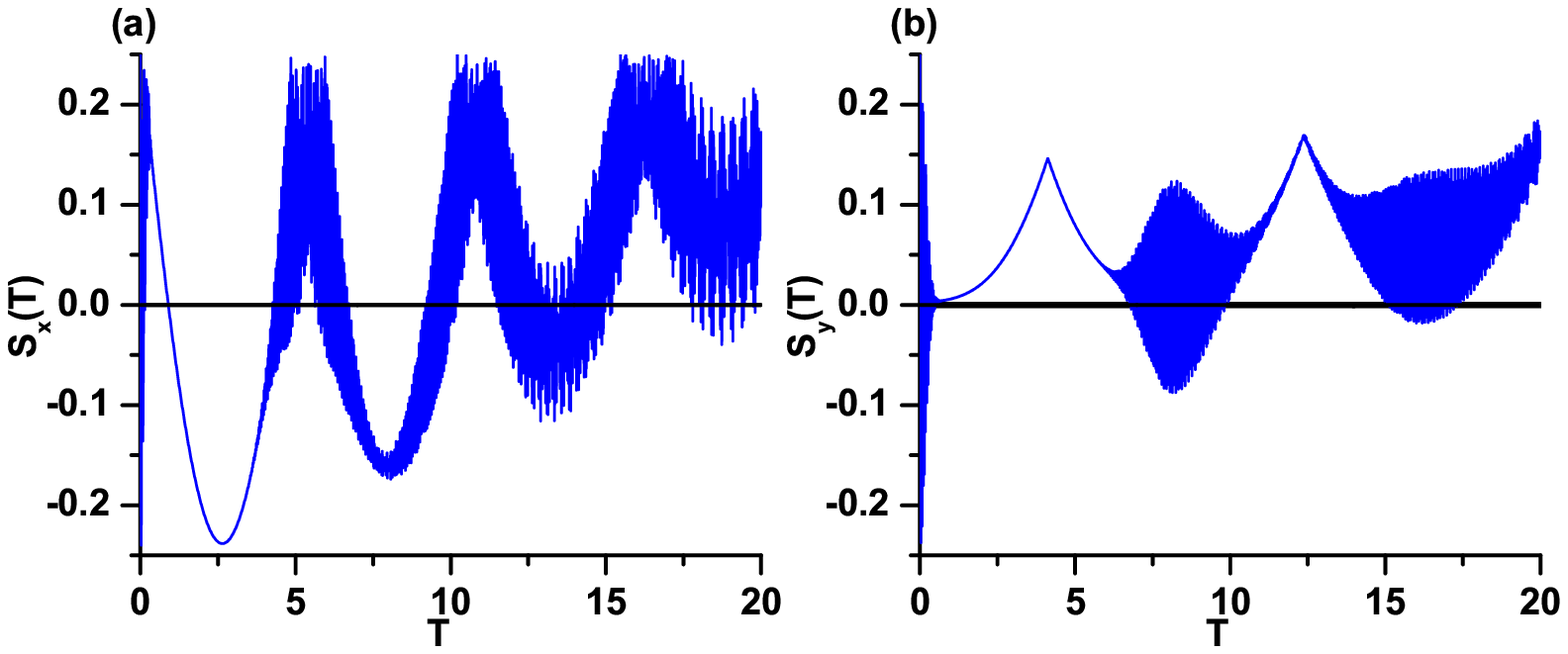}
\caption{ The variances squeezing as indicated
against the scaled time $T$  for $(j,\protect\theta)=(150,\protect\pi %
/3)$ with  $\bar{\Delta}=0$ (a) and  $300$ (b). Figure (a) is
given to the case $(x, z)$, while (b) for the case $(y, z)$. The
straight line represents the squeezing bound.
 }
\end{figure}
%%%%%%%%%%%%%%%%%%%%%%%%%%%%%%%%%%%%%%%%%%%%%%%%%%%%%%%%%%%%%
%%%%%%%%%%%%%%%%%%%%%%%%%%%%%%%%%%%%%%%%%%%%%%%%%%%%%%%%%%%%

\begin{figure}[tbh]
\centering
\includegraphics[width=20pc,height=15pc]{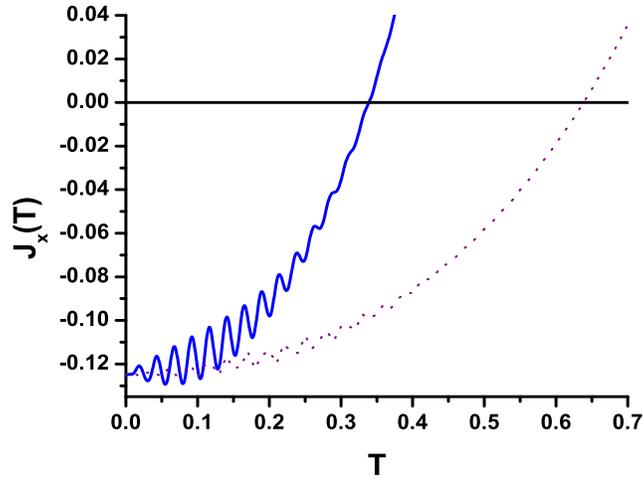}
\caption{The variance squeezing $J_{x} $ of the angular
momentum against the scaled time $T$ where $\left( j,\bar{\Delta},\protect%
\theta \right) =\left( 150,0,\protect\pi /3\right) $\ solid curve and $%
\left( j,\bar{\Delta},\protect\theta \right) =\left( 150,300,\protect\pi %
/3\right) $ dashed curve. The straight line represents the
squeezing bound. }
\end{figure}

Now we turn our attention to discuss the  squeezing related to the
$\widetilde{j}$ system. For the case $(x,y)$   we have
 $\hat{A}=\hat{J}%
_{x},\hat{B}=\hat{J}_{y}$ and $\hat{C}=\hat{J}_{z}$ and the
information is shown in Fig. 6. Generally, we have found that
squeezing occurs only in the $x$-component. Precisely,  the CSS
exhibits squeezing for certain values of $\theta$. This initial
squeezing gradually decreases and vanishes as the interaction
proceeds. Squeezing in the off-resonance case  can survive for
period of interaction time longer than that of the resonance one.
This is obvious when we compare the solid curve to the dashed
curve in Fig. 6. The occurrence of the nonclassical effects in the
$\widetilde{j}$ system basically depends on the initial state of
the $\widetilde{j}$ system
\cite{physica1,physica2,physica3,physica4,physr4}.

\section{Entanglement}

In this section, we investigate the entanglement in the system.
Precisely, we demonstrate two type of entanglement; namely
central-atom-surroundings entanglement and entanglement among the
surrounded spins. Before going into the details there is an
important remark we would like to address here. In somehow the
interaction mechanism in the system (2) is quite similar to that
of the JCM \cite{1}. This is quite obvious  if we interchange the
bosonic operators by the angular momentum ones, i.e.
$(\hat{J}_-,\hat{J}_+,\hat{J}_z)\longrightarrow
(\hat{a},\hat{a}^\dagger,\hat{a}^\dagger \hat{a})$.  We exploit
this fact to use the linear entropy  for quantifying the
entanglement between the $\widetilde{s}$ and $\widetilde{j}$
systems  \cite{c,f}. It is worth mentioning that such type of
entanglement is completely different from that discussed in
\cite{physica1}, where the authors have studied the entanglement
between two central atoms after tracing over the variables of the
surroundings. Now, we start the discussion with the
$\widetilde{s}-\widetilde{j}$ entanglement. For this purpose we
write down the definition of the linear entropy as  \cite{c,f}:
\begin{equation}
\eta \left( T\right) =\frac{1}{2}\left[ 1-\xi ^{2}\left( T\right)
\right], \label{5.1}
\end{equation}%
where $\xi \left( T\right) $ is the well known a Bloch sphere
radius, which has the form:
\begin{equation}
\xi ^{2}\left( T\right) =\langle \hat{S}_{x}\left( T\right)
\rangle ^{2}+\langle \hat{S}_{y}\left( T\right) \rangle
^{2}+\langle \hat{S}_{z}( T) \rangle ^{2}. \label{5.2}
\end{equation}
%%%%%%%%%%%%%%%%%%%%%%%%%%%%%%%%%%%%%%%%%%%%%%%%
The Bloch sphere is a tool in quantum optics, where the simple
qubit state is successfully represented, up to an overall phase,
by a point on the sphere, whose coordinates are the expectation
values of the atomic set operators of the system. This means that
entanglement is strictly related to the behavior of observable of
clear physical meaning. The function $\eta ( T) $ ranges between
$0$ for disentangled bipartite and $1/2$ for maximally entangled
ones. In Figs. (7) we have plotted the function $\eta ( T) $
against the scaled time $T$ for given values of the interaction
parameters. From Fig. 7(a), i.e. $\theta=\pi/2$, as the
interaction time increases the decoupled bipartite at $T=0$
becomes entangled. The entanglement gets stronger as the
interaction proceeds. Also the function shows rapid fluctuations
during the whole duration  of the interaction. The bipartite is
periodically disentangled with period  $\tau \simeq 480$. This in
a good agreement with the behavior of the atomic inversion, (see
Fig. 1(a)). Now we extend our discussion to the case $\theta =\pi
/3$ (see Figs. 7(b)-(c)). For the resonance case, the bipartite
exhibits
 maximal entanglement  after switching on  the interaction.
Afterward, the function $\eta \left( T\right) $ rebounds and
decreases its value below $0.4$ around $T\simeq 2.5$ showing
partial entanglement, see Fig. 7(b). As the interaction is going
on the bipartite exhibits long-lived entanglement. In some sense
the
 behavior of Fig. 7(b) is quite similar to that of the JCM
\cite{f}. Information about the off-resonance case is shown in
Fig. 7(c). From this figure
 we can observe a considerable reduction in the entanglement
 compared to the previous case. Also this figure
 exhibits RCP  as that of the  corresponding atomic
 inversion. The comparison among Figs. 7(a)--(c) shows that the values of
 the parameters $\theta$ and  $\bar{\Delta}$ play the significant
 role in the occurrence of the RCP.

\begin{figure}[tbh]
\centering
\includegraphics[width=35.pc,height=15pc]{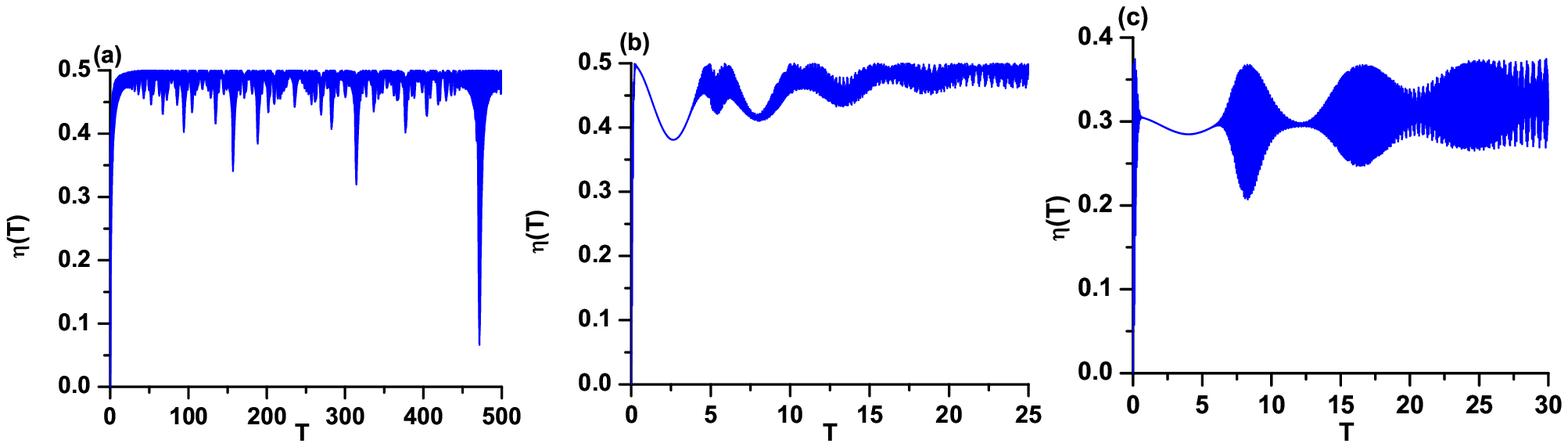}
\caption{Linear entropy $\protect\eta \left( T\right) $ against
the scaled time $T$  for $( j,\bar{\Delta},\protect\theta ) =(
150,0,\protect\pi /2) $ (a),  $( 150,0,\protect\pi /3)$ (b), and
$( 150,300,\protect\pi /3)$ (c).}
\end{figure}

Now, we close this section by investigating the entanglement
 among the spins in the $\widetilde{j}$ system.  The entanglement among
atoms in the $N$-particle system  can be amounted by spin
squeezing concept  as \cite{choi,fs}:

\begin{equation}
\bar{\zeta}( T) =\frac{N\langle \left( \Delta
\hat{J}_{x}(T)\right)
^{2}\rangle }{\langle \hat{J}_{y}( T) \rangle ^{2}+\langle \hat{J}%
_{z}( T) \rangle ^{2}},  \label{5.3}
\end{equation}%
where $N=2J$. The system shows disentanglement (maximal
entanglement) when $\bar{\zeta}\left( T\right) =1 \quad (0)$. For
the sake of convenience we reformulate the above equation
(\ref{5.3}) as:
\begin{equation}
\zeta ( T) =\langle \left( \Delta \hat{J}_{x}(T)\right) ^{2}\rangle -%
\frac{1}{N}\left[ \langle \hat{J}_{y}( T) \rangle ^{2}+\langle
\hat{J}_{z}( T) \rangle ^{2}\right] . \label{5.4}
\end{equation}%
\begin{figure}[tbh]
\centering
\includegraphics[width=25pc,height=20pc]{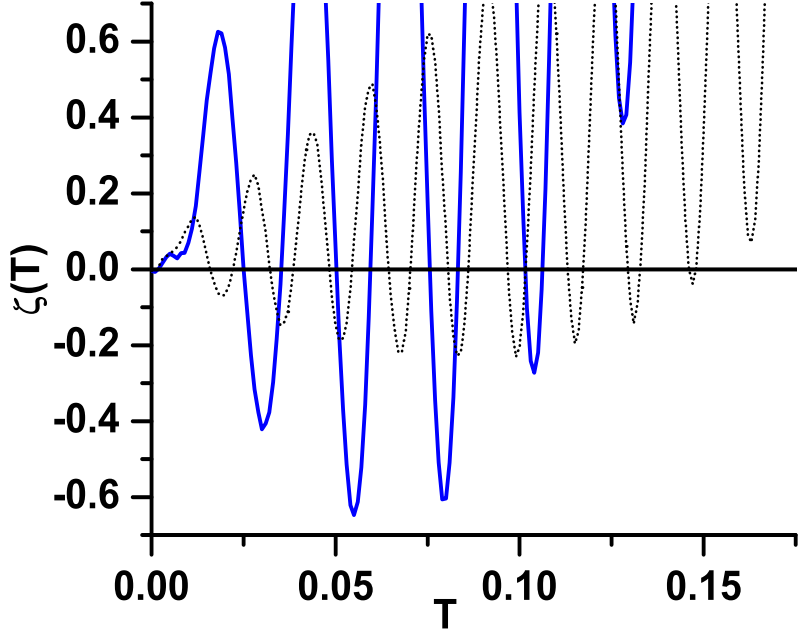}
\caption{Entanglement parameter $\protect\zeta \left( T\right) $
of the
angular momentum system in the $x$-direction aganist the scaled time $T$ \ for $%
\left( j,\bar{\Delta},\protect\theta \right) =\left( 150,0,\protect\pi %
/3\right) $ solid curve and $\left( 150,300,\protect\pi /3\right)
$ dash curve.}
\end{figure}
In this case
 the  limitations become  $\zeta ( T)
=-1$ for maximal entanglement and $0$ for disentanglement. Now we
use the formula (\ref{5.4}) to examine the  behavior of our
system. The calculation of different quantities in (\ref{5.4}) is
already given in section 2.  In Fig. 8 we plot $\zeta ( T)$  for
$( j,\bar{\Delta},\theta ) =( 150,0,\pi /3) $ (solid curve) and
 $( 150,300,\pi /3) $ (dashed curve). From this figure we can
observe that the entanglement instantaneously  occurs among the
spins only for a short duration of interaction time. The amounts
of entanglement in the resonance case are stronger than those in
the off-resonance one. Nevertheless, the increase in
$\bar{\Delta}$ results in the increase in oscillation frequency of
the $\zeta ( T)$. This indicates a potential for the quantum
control of entanglement properties of the system via externally
adjustable parameters. The comparison between the Fig. 6 and the
Fig. 8 confirms the fact that the observed squeezing is not always
accompanied by an increase in the quantum entanglement among the
spins \cite{choi}. Precisely, quantum squeezing does not
necessarily imply entanglement.

\section{Conclusion}

In this paper, we have studied the quantum properties of  the $XX$
central spin model. We have mange to solve the Heisenberg
equations of motion for the Pauli spin operators and the angular
momentum ones. We have considered  that the $\widetilde{s}$ and
$\widetilde{j}$ systems are initially prepared in the excited
state and the CSS, respectively. We have investigated the atomic
inversion, the variance squeezing and the entanglement. The
results can be summarized as follows. The atomic inversion can
exhibit RCP  for certain values of the interaction parameters.
Precisely, the
 occurrence of the RCP is sensitive not only to the smoothness of
 the photon distribution but also to its location of
 the symmetrization center  in the photon-number domain. Squeezing is
 available only in the $x$ and $y$ components of the Puali spin
 operators. Also controlling the value of $\bar{\Delta}$ can lead
 to the occurrence of the RCP in the atomic squeezing. Squeezing can occur in the
 $\widetilde{j}$ system only when CSS is initially squeezed.  The
 entanglement between the bipartite  $\widetilde{s}$ and
 $\widetilde{j}$ is sensitive to the value of $\theta$. For
 instance, when $\theta=\pi/2$ the bipartite becomes  periodically
 disentangled.  Increasing the value of $\bar{\Delta}$ leads to
 decreasing the entanglement in the bipartite.  Entanglement among
 the spins of the $\widetilde{j}$ system occurs for short duration
 of the interaction time. Also quantum squeezing does not
necessarily imply entanglement.

\textbf{Acknowledgement: }

One of us (M.S.A.) is grateful for the financial support from the
project Math 2010/32 of the Research Centre, College of Science,
King Saud University.

\end{document}